\documentclass[12pt]{article}
\usepackage{amsmath,amssymb,mathtools,amsfonts,epsfig,graphicx,euscript}
 \usepackage{color}
 \usepackage{etoolbox}
 \usepackage{xcolor}
\usepackage{enumitem}
\usepackage{multirow}
\usepackage{float}
\usepackage{appendix}
\begin{document}
\begin{titlepage}
\thispagestyle{empty}
IPM/P-2020/024
\vspace{1cm}
\begin{center}
\font\titlerm=cmr10 scaled\magstep4
\font\titlei=cmmi10
scaled\magstep4 \font\titleis=cmmi7 scaled\magstep4 {\Large{\textbf{Gravitational waves of a first-order QCD phase transition at finite coupling from holography}
\\}}
\setcounter{footnote}{0}
\vspace{1.5cm} 
\noindent{{M. Ahmadvand$^a$ \footnote{e-mail:ahmadvand@ipm.ir} K. Bitaghsir Fadafan$^b$ \footnote{e-mail:bitaghsir@shahroodut.ac.ir} and S. Rezapour$^b$ \footnote{e-mail:S.Rezapour@shahroodut.ac.ir}
		}}\\
\vspace{0.2cm}

{\it $^a$School of Physics, Institute for Research in Fundamental Sciences (IPM), P. O. Box 19395-5531, Tehran, Iran\\}
{\it $^b$Physics Faculty, Shahrood University of Technology, P.O.Box 3619995161 Shahrood, Iran\\}

\vspace*{.4cm}
\end{center}

\vskip 2em
\setcounter{footnote}{0}
\begin{abstract}
We consider a holographic study of coupling dependent gravitational waves produced by the cosmic first-order QCD phase transition at finite baryochemical potential. In the dual description, the first-order QCD phase transition corresponds to the first-order Hawking-Page phase transition in Gauss-Bonnet gravity. At intermediate coupling, we obtain key quantities characterizing the gravitational wave energy density spectrum. We then find that the gravitational waves might be detected for sufficiently large Gauss-Bonnet coupling, while sound waves play an important role in the spectrum. We also consider a supercooling scenario during the QCD phase transition and show that the gravitational waves generated during this period can be detected by pulsar timing array experiments.

\end{abstract}
\end{titlepage}

\baselineskip=1.3\baselineskip	
\section{Introduction}
Heavy ion collision experiments in the Large Hadron Collider (LHC) and Relativistic Heavy Ion Collisions (RHIC) study the physics of strongly interacting matter, including the key features of quark-gluon plasma and the phase structure of de/confinement and chiral Phase Transitions (PTs), which are predicted by Quantum Chromodynamics (QCD). In fact, these colliders recreate the condition which has inversely occurred in the early stages of evolution of the Universe.

Around $ 10^{-6}$ seconds after the big bang, the strongly interacting quarks and gluons were confined and entered into the hadronic phase. At the energy scales which this de/confinement PT occurs, the theory is strongly coupled and we cannot use perturbative methods in QCD. At high temperatures, for static quarks, lattice QCD shows the PT is first-order \cite{Lucini:2012wq}, while for three light quarks at small baryochemical potential, it is a crossover \cite{Bhattacharya:2014ara}. However, there are possibilities under which the PT becomes first-order in the early Universe. A strong mechanism for baryogenesis may lead to a large amount of baryochemical potential \cite{Dine:2003ax}. Moreover, for large lepton asymmetry, finite baryochemical potential can be produced \cite{Schwarz:2009ii}. This finite baryochemical potential results in a first-order QCD PT \cite{Schwarz:2009ii,Schettler:2010dp}. 

Since cosmological first-order PTs can be regarded as possible sources of Gravitational waves (GWs) \cite{Witten:1984rs}, the study of these PTs becomes more important. On the other hand, recent progresses and also future plans for detection of GWs have augmented this motivation. Cosmological PTs, especially electroweak and QCD PTs, under some conditions can be first-order \cite{Schwarz:2009ii,Abedi:2019msi}. During a first-order PT, bubbles, which separate two phases of the PT, nucleate and expand. The expansion of these bubbles in the plasma and eventually their collision with each other lead to the GW generation in the space-time \cite{Kosowsky:1992rz,Kamionkowski:1993fg,Kosowsky:2001xp,Hindmarsh:2015qta}. 

Recently, we have initiated a holographic study of GWs generated during the cosmic first-order QCD phase transition in \cite{Ahmadvand:20172} and continued our study by considering the effect of non-zero baryochemical potential in the plasma  \cite{Ahmadvand:2017tue}. \footnote[1]{In this holographic approach of GW study, for a technicolor model see \cite{Chen:2017cyc}.} Within AdS/QCD models, we considered large $N_c$ field theories at infinitely strong 't Hooft coupling, $\lambda$. In this holographic approach, there is a correspondence between the Hawking-Page PT and de/confinement PT \cite{Witten:1998zw,Herzog:2006ra}. Using this correspondence, we studied the GW energy density spectrum at the infinite coupling. However, one expects that the PT occurs at the intermediate coupling constant. Then, it is desirable to study $1/\lambda$ series corrections around the infinite $\lambda$. From holography, one needs to study higher derivative terms in the action. In this work, we consider Gauss-Bonnet gravity as curvature-squared corrections. One should notice that leading order corrections to type IIB string theory are proportional to $\alpha'^3 $, where $\alpha' $ is the string tension, and the dual field theory is a  $\mathcal{N}=4$ supersymmetric Yang-Mills theory. Although a precise dual theory to the Gauss-Bonnet gravity is unknown, one can study the expectation value of dual operators in the framework of dualities. As a result, this work helps to understand the effect of finite coupling corrections on the holographic study of GWs generated during the first order QCD PT at non-zero baryochemical potential. Actually, it helps to construct these holographic models more accurately. In this framework, through a soft wall model, we study the first-order Hawking-Page PT, between thermal charged AdS and the black hole solution. Using the behavior of the string of a quark-antiquark, for different values of Gauss-Bonnet coupling, we obtain the critical temperature and baryochemical potential at this point. To characterize the GW energy density spectrum, we calculate other key quantities, such as the vacuum energy density of the PT and the temperature at which GWs are generated. 

In the case of high baryochemical potential provided by a strong baryogenesis mechanism, a short period of supercooling might take place during the QCD PT and  hence the baryon density could be diluted. In this paper, we take into account such a scenario and calculate the required quantities to obtain the GW spectrum, within our model. We indicate these signals can be in the sensitivity range of future-planned GW detectors, i.e., SKA and IPTA.

The paper is organized as follows. In Section 2, we introduce the gravitational background and study the PT dynamics. In Section 3, we compute the GW spectrum for different cases. We conclude in Section 4.  

\section{ Gravity background and de/confinement phase transition  }

First we study the gravity background. The Gauss-Bonnet term contribute in the bulk only for dimensions higher than four. In d = 4, the Gauss-Bonnet contribution is a  topological term and has no dynamics. \footnote[2]{Recently  a  non-trivial  four-dimensional  Einstein-Gauss-Bonnet theory of gravity has been proposed in \cite{Glavan:2019inb}.} This theory is an example of Lovelock backgrounds where do not have instability problems. They are interesting theories to study non-perturbative effects with considering higher derivative corrections, for example see \cite{Grozdanov:2016fkt} studying transport coefficients and quasinormal spectra in the  Gauss-Bonnet holographic fluid. The most important effect is the violation of the bound on the ratio of shear viscosity, $\eta$, to entropy density, $s$, which is expressed as \cite{Brigante:2007nu}
\begin{equation}
\frac{\eta}{s}=\frac{1}{4 \pi}\left(1-4\lambda_{\mathrm{GB}}\right)
\end{equation}
where $\lambda_{\mathrm{GB}} $ is the Gauss-Bonnet coupling with dimension $(length)^2$. One finds that the gravity background with $\lambda_{\mathrm{GB}}<0$ is dual to the field theory with $\eta/s$ larger than $1/4\pi$ while $\lambda_{\mathrm{GB}}>0$ corresponds to the smaller ratio. On the other hand, it is known that lower (higher) viscosities can be intuitively considered as more (less) strongly coupled field theories \cite{Noronha:2009ia}. In the following, we will conclude and interpret our results based on these relations.

As for $\lambda_{\mathrm{GB}} $ values, we should notice some constraints. For $ \lambda_{\mathrm{GB}}<  \frac{1}{4}$, there is no Conformal Field Theory (CFT) and also vacuum AdS solution \cite{Brigante:2008gz}. Moreover, the positivity constraint of the energy in CFTs and avoiding causality violation constrain the value of $ \lambda_{\mathrm{GB}}$ as \cite{deBoer:2009pn,Camanho:2009vw,Camanho:2009hu}
\begin{equation}
-7/36 < \lambda_{\mathrm{GB}} \leq 9/100.
\label{b1}
\end{equation}
We respect this bound though from a phenomenological point of view, we also consider $ \lambda_{\mathrm{GB}}=0.12$ in our calculations.

The Euclidean action for Einstein-Maxwell theory with the Gauss-Bonnet coupling in five dimensions can be written down as \cite{Sachan:2013zza,Bhatnagar:2017qyb}  \footnote[3]{One finds surface terms and counterterms in \cite{Cvetic:2001bk} and \cite{Brihaye:2008xu}.}
\begin{equation}
S =-  \int{ d^5 x \sqrt{g}\, e^ \phi \left [ \frac{1}{2  \kappa^2} \Big(\mathcal{R}-2 \Lambda+\lambda_{\mathrm{GB}} \mathcal{R}_{\mathrm{GB}}\Big)-\frac{1}{4 g^2}F^2 \right] }
\label{eq1}
\end{equation}
where the Gauss-Bonnet term is  $\mathcal{R}_{\mathrm{GB}}=\mathcal{R}^2- 4 \mathcal{R}_{\mathrm{\mu \nu}}\mathcal{R}^{\mathrm{\mu \nu}}+\mathcal{R}^{\mathrm{\mu \nu \rho \sigma}}\mathcal{R}_{\mathrm{\mu \nu \rho \sigma}} $, $ F^2 =F_{\mathrm{\mu \nu}} F^{\mathrm{\mu \nu}}$ is the intensity of the Maxwell field and $\kappa^2= 8\pi G_5$. The field $\phi $ is a dilaton field expressed as $\phi(z)= -cz^2$ where  $\sqrt{c}=388\,\mathrm{MeV} $ is determined due to the calculation of the lightest $\rho$ meson mass \cite{Karch:2006pv}. This field is non-dynamical, hence the equations of motion remain unchanged. The AdS/QCD correspondence relates the five-dimensional gravitational constant $\kappa$ and five-dimensional coupling constant $g$ to the number of color $N_c$ and flavor $N_f$ as \cite{Sin:2007ze}
\begin{equation}
\frac{1}{2\kappa^2}= \frac{{N_c}^2}{8 \pi^2}\,, \hspace{1.cm} \hspace{1.cm}  \frac{1}{2  g^2}= \frac{N_c N_f}{8 \pi^2}. 
\label{eq13}
\end{equation}
Next we study the phase structure of QCD with Gauss-Bonnet corrections in a soft wall model whose dilaton field is introduced in Eq.\ (\ref{eq1}). The quark-qluon plasma, as the deconfinement phase, with high density and temperature corresponds to the charged AdS black hole in the Gauss-Bonnet background. From Eq.\ (\ref{eq1}), the black hole solution is given by 
\begin{equation}
ds^2 = \frac{1}{z^2} \left(A^2 f(z) dt^2+ \frac{ dz^2}{f(z)}+ \sum\limits_{\mathrm{i=1}}^3 {\mathrm{dx^i}}^2 \right)
\label{eq s}
\end{equation}
where $ A^2 =\frac{1}{2}( \sqrt{1-8\lambda_{\mathrm{GB}}}+1)$ and the blackening metric function is given by

\begin{equation}
f(z) = \frac{1}{4 \lambda_{\mathrm{GB}}} \left (1- \sqrt{1-8\lambda_{\mathrm{GB}} (1-m z^4+q^2 z^6)}\right ).
\label{eq f}
\end{equation}
Here $z$ is the radial coordinate and the boundary space coordinates are denoted as $\mathrm{x^i}$. The boundary is also located at $z=0$. Moreover, one finds the solution of gauge equation as
\begin{equation}
\mathcal{A}_t(z) = i (\mu - Q  z^2)
\label{eq2}
\end{equation}
where $\mathcal{A}_t $ is the time component of the bulk gauge field and $\mu$ is the baryochemical potential. Since solutions of $\mathcal{A}_t(z)$ are regular at the horizon, $z_h$, we impose Dirichlet boundary condition at the horizon as $\mathcal{A}_t(z_h) = 0$, leading to the relation $Q = \mu/z_h^2$. One should notice that $Q$ is the quark number density which is related to the AdS space charge denoted by $q$ as
\begin{equation}
q =\sqrt{ \frac{2}{3}}  \frac{\kappa Q }{g A}.
\label{eq5}
\end{equation}
According to the condition that the metric function is zero on the horizon, we obtain  the black hole mass, $m$, as
\begin{equation}
m= \frac{1}{{z_h}^4}+q^2 {z_h}^2. 
\label{eqY}
\end{equation}
Also, the Hawking temperature of the black hole with Gauss-Bonnet corrections is given by
\begin{equation}
T= \frac{A f^{\prime}(z_h)}{4 \pi}= \frac{A}{\pi z_h} \left(1- \frac{1}{2}q^2 {z_h}^6\right).
\label{eq9}
\end{equation}
Using this relation, the positive solution for the horizon can be obtained as
\begin{equation}
z_h=\frac{3}{2}  \frac{g^2 A}{\kappa^2  \mu^2 }  \left ( \sqrt{\frac{4}{3}  \frac{\kappa^2  \mu^2 }{g^2}+ \pi^2  T^2- \pi T}\,\right ). 
\label{eq12}
\end{equation}
To find the free energy of the system in the deconfinement phase, we should take the integrals in Eq.\ (\ref{eq1}). However, this cannot be done analytically. Thus, we expand the action up to the first-order in $ \lambda_{\mathrm{GB}}$
\begin{align}
S& =-\frac{A V_3}{ k^2} \int_0^{ T^{-1} }{ dt}  \int_\varepsilon^{z_h}{dz\frac{e^{-cz^2}}{z^5}} \nonumber\\
&\Big [2 q^2 z^6-4+ \lambda_{\mathrm{GB}} \Big(24  m^2 z^8- 120 m q^2 z^{10}+112 q^4 z^{12} +8 q^2 z^6+ 40\Big)\Big]+\mathcal{O}( \lambda_{\mathrm{GB}}^2)
\label{eq15}
\end{align}
where $V_3$ is the three-dimensional volume and $\varepsilon$  is $UV$ cutoff. Taking  the limit  $\varepsilon \rightarrow 0$, one finds the free energy density as follows:
\begin{align}
F_{\mathrm{BH}} &=-\frac{A}{\kappa^2} \left [e^{-c z_h^2} \left( \frac{c z_h^2-1}{z_h^4} + \frac{q^2}{c}\right) + c^2 Ei(- c z_h^2)-\frac{q^2}{c}+\frac{m}{2} \right. \nonumber\\
&\left. + \lambda_{\mathrm{GB}}\left [ e^{-c z_h^2} \left( \frac{10 (c z_h^2-1)}{z_h^4} - \frac{4 q^2}{c}- 12 m^2  \left (\frac{1}{c^2}+\frac{z_h^2}{c}\right)\right) \right.  \right. \nonumber\\
&\left.\left. + 60 mq^2  \left(\frac{2}{c^3}+\frac{2 z_h^2}{c^2}+\frac{z_h^4}{c}\right)-56 q^4  \left(\frac{6}{c^4}+\frac{6 z_h^2}{c^3}+\frac{3 z_h^4}{c^2}+ \frac{z_h^6}{c}\right) \right.  \right. \nonumber\\
&\left.\left.+ \left(\frac{12 m^2}{c^2}-\frac{120 mq^2}{c^3}+\frac{336 q^4}{c^4}\right)- 4 m+10 c^2 Ei(-c z_h^2)\right]\right]+\mathcal{O}( \lambda_{\mathrm{GB}}^2)
\label{eqBH}
\end{align}
where the exponential integral is given by $Ei(x)=-\int _{-x}^{\infty} dt\, e^{-t}/t$.\\

Based on the holography, the confined phase corresponds to the metric for thermally charged AdS with Gauss-Bonnet terms, which is given by 
\begin{equation}
ds^2 = \frac{1}{z^2} \left (A^2 f_1(z) dt^2+ \frac{ dz^2}{ f_1(z)}+ \sum\limits_{\mathrm{i=1}}^3 {\mathrm{dx^i}}^2 \right ).
\label{eq10}
\end{equation}
The metric function $f_1(z)$ is defined as 
\begin{equation}
f_1(z) = \frac{1}{4 \lambda_{\mathrm{GB}}} \left (1- \sqrt{1-8\lambda_{\mathrm{GB}} (1+q_1^2 z^6)}\right ). 
\label{eq4}
\end{equation}
In this phase
\begin{equation}
Q_1 =3 c \mu/2,~~~~~~~~~~~q_1 = \sqrt{\frac{3}{2}} \frac{\kappa}{g}\frac{c\mu}{A}.
\end{equation}
Furthermore, the free energy density for thermally charged AdS up to first-order in  $\lambda_{\mathrm{GB}}$  is given by
\begin{align}
F_{\mathrm{th}}= -\frac{A}{\kappa^2} \left [\frac{q_1^2}{c }+ 4 \lambda_{\mathrm{GB}} \left( \frac{ q_1^2}{c}- \frac{84 q_1^4}{c^4} \right)\right]+\mathcal{O}( \lambda_{\mathrm{GB}}^2).
\label{eqth}
\end{align}
Having introduced the deconfined and confined phases, we can study the transition between these phases, using AdS/CFT. Based on this duality, this PT corresponds to the Hawking-Page PT. Also, the Hawking-Page PT in this background has been studied in \cite{Sachan:2013zza}.\\

From Eq.\ (\ref{eqBH}) and Eq.\ (\ref{eqth}), the free energy density difference is obtained as
\begin{align} 
\Delta F & =F_{\mathrm{BH}} - F_{\mathrm{th}} =  \nonumber \\
& \frac{A}{\kappa^2}\left [e^{-c z_h^2}\left ( \frac{c z_h^2-1}{z_h^4} + \frac{q^2}{c}\right) + c^2 Ei(-c z_h^2)+  \frac{q_1^2 - q^2}{c}+\frac{m}{2}\right. \nonumber\\
&\left.+\lambda_{\mathrm{GB}}\left [e^{-cz_h^2}\left ( \frac{10 (c z_h^2-1)}{z_h^4}+ \frac{4 q^2}{c}+12m^2\left (\frac{1}{c^2}+\frac{z_h^2}{c} \right)\right. \right. \right.\nonumber\\
&\left.\left.\left.- 60 m q_1^2\left (\frac{2}{c^3}+\frac{2 z_h^2}{c^2}+\frac{z_h^4}{c} \right)+56 q_1^4\left (\frac{6}{c^4}+\frac{6z_h^2}{c^3}+\frac{3 z_h^4}{c^2}+\frac{z_h^6}{c} \right) \right)\right.\right. \nonumber\\
&\left.\left.-10 c^2 Ei(-c z_h^2)+ 4 \left( \frac{84 q_1^4}{c^4}+  \frac{q_1^2 - q^2}{c}- m \right)\right.\right. \nonumber\\
&\left.\left.-\left( \frac{12 m^2}{c^2} -\frac{120 mq^2}{c^3}+\frac{336 q^4}{c^4}\right) \right]\right]+\mathcal{O}( \lambda_{\mathrm{GB}}^2).
\label{delfree}
\end{align}
At the critical temperature, $T_c$, the free energies become degenerate so that $ \Delta F=0$. Using this equation, we find a relation between $\mu$ and $z_h$. To specify these parameters at the transition, we can use the expectation value of Polyakov loop, as the order parameter, which is related to the heavy quark-antiquark potential. (For  details, see \cite{Ahmadvand:2017tue}.) Then, we consider a U-shaped open sting whose end points on the boundary space-time are quark anti-quark pairs. In fact, It can be regarded as a meson in the field theory side. In the confined phase, a meson is stable and the U-shaped string reaches a maximum value at $z=z_*$ which always is located behind the wall, while in the deconfined phase the black hole absorbs the string and the maximum can reaches the horizon. Thereby the meson melts in the medium. Higher derivative corrections on the meson melting have been studied in \cite{AliAkbari:2009pf,Fadafan:2011gm,Fadafan:2012qy,Fadafan:2013bva,Fadafan:2013coa}.

We consider the meson in the decofined phase and numerically finds the chemical potential for each $\lambda_{\mathrm{GB}}$ so that the U-shaped string in the bulk reaches the $z_ *$. This procedure has been done in \cite{Ahmadvand:2017tue}. Here, we summarize the steps to get some main quantities that we need to calculate the possible GW spectrum.
\begin{itemize}
	\item Using the U-shaped string in the confined phase, we fix the chemical potential so that the maximum depth that the string falls through the bulk touches the wall $ z =  \frac{1}{\sqrt{c}}$.  
	
	\item The first-order PT occurs when $\Delta  F = 0$. Therefore, by fixing the chemical potential, the value of $ z_h$ can be obtained.
	
	\item Putting $\mu$ and $ z_*$ values in Eq.\ (\ref{eq9}), the value of $T_c$ is determined.
	
\end{itemize}

Using the above steps, we show the chemical potential and critical temperature for different values of the Gauss-Bonnet coupling in Table \ref{t1}. Although $\lambda_{\mathrm{GB}}=0.12$ is not in the range of Eq.\ (\ref{b1}), it is interesting to explore phenomenologically larger values of $\lambda_{\mathrm{GB}}$.

\begin{table}[H]\label{t1}
	\begin{center}
		\begin{tabular}{| c| c| c| c|} 
			\hline
			$  N_ f $  & $\lambda_{\mathrm{GB}}$  & $ \mu [\mathrm{MeV}]$&$ T_c[\mathrm{MeV}] $  \\
			\hline\hline
			\multirow{7}{*}{2} &0.12 &35& 61.7 \\ 
			\cline{2-4}
                                &0.06 &72 &122.9 \\
			\cline{2-4}
			& 0.03 & 80 &156.1 \\
			\cline{2-4}
			&0 &86 &191.6\\
			\cline{2-4}
			&-0.03 &90 &226.7\\
			\cline{2-4}
			&-0.06 &95 &260.4\\
			\cline{2-4}
			&-0.12 &102 &324\\
			\hline
		\end{tabular}
		\caption{The values of the chemical potential and critical temperature for different values of the Gauss-Bonnet coupling, $\lambda_{\mathrm{GB}}$, and $ N_ f= 2$ are listed. }
		\label{t1}
	\end{center}
\end{table}
In the next section, using the above values of these parameters, we proceed to study the GW spectrum in the presence of Gauss-Bonnet corrections.

\section{Gravitational Wave Spectrum}

Detection of GWs has provided a promising way to probe early Universe events. One of the sources producing these GWs is a cosmological first-order PT. For first-order PTs, during bubble nucleation, three sources of GWs have been proposed: collision of the bubbles, sound waves and turbulent motion of the fluid. The sum of these three sources contributing to the GW energy density spectrum can be added together as follows
\begin{equation}
h^2 \Omega(f)\simeq h^2 \Omega_{\mathrm{col}}(f)+h^2 \Omega_{\mathrm{sw}}(f)+h^2 \Omega_{\mathrm{tu}}(f)
\end{equation}
where $h$ is the present Hubble parameter $H_0$ in units of 100 km $\mathrm{sec}^{-1}\mathrm{Mpc}^{-1} $. The contribution of each GW source is characterized by important PT parameters, including the vacuum energy released during the PT, duration of the PT, which are calculated at the time of GW generation, and bubble wall velocity.

The parameter associated with the vacuum energy is given by
\begin{equation}
\alpha= \frac{\epsilon _*}{\rho _R( T_*)}
\end{equation}
where $ \alpha $ is the ratio of the vacuum energy density to the radiation energy density, $ T_*$ denotes the temperature at which GWs are produced and
\begin{equation}
\epsilon _*=\left(\Delta F(T)-T\frac{d\Delta F(T)}{dT}\right)\Bigg|_{T=T_*},~~~~~~~~\rho _R( T_*)=  \frac{\pi ^2}{30} g_* T_*^4.
\end{equation}
Another important quantity is $\beta/H_* $, where $\beta ^{-1}$ is approximately the duration of the PT, $H_*=1.66\sqrt{g_*}T^2_*/M_p$ is the Hubble parameter calculated at $T_*$, $ g_*$ is the number of effective relativistic degrees of freedom in the plasma at this temperature and $ M_p$ is the Planck mass. The dependence of energy density and peak frequency of each source on these PT parameters can be found in Appendix \ref{gwsources}. 

In this section, we obtain the PT characteristics, based on the model that we introduced in the previous section and then calculate the GW spectrum of the first-order QCD PT. Moreover, we examine whether the generated GWs fall within the sensitivity range of pulsar timing experiments.

To obtain $T_*$, we use the following procedure. The duration of the PT can be well approximated by \cite{Leitao:2012tx}
\begin{equation}
\Delta t=t_*-t_c=\frac{3}{\beta}\log (\frac{\beta}{H}).
\end{equation}
Furthermore, using $dT/dt \simeq -HT$, we obtain
\begin{equation}
\Delta t\simeq M_p\frac{T_c^2-T_*^2}{2T_c^2T_*^2}.
\end{equation}
Therefore, from the above relations we obtain
\begin{equation}
\frac{\beta/H_*}{3\log(\beta/H_*)}=\frac{2T_c^2}{T_c^2-T_*^2}.
\end{equation}
Finally, from the result computed in the previous section for $ T_c$ and also by fixing $\beta/H_*=100$ \cite{Caprini:2015zlo} for the PT, we can find $ T_*$.\footnote[4]{In our previous works \cite{Ahmadvand:20172,Ahmadvand:2017tue}, we considered $ T_*\simeq T_c$.}

To determine the GW spectrum, $\alpha$ should also be attained. Thus, taking into account the result obtained for $\Delta F $, Eq.\ (\ref{delfree}), we can calculate this ratio at $ T_*$. In addition, there is a critical value of $\alpha$, denoted by $\alpha_{\infty} $, which can also specify the speed of bubble walls:
\begin{equation}
\alpha_{\infty}=\frac{30}{24 \pi ^2}\frac{\sum_i c_i \Delta m_i ^2}{g_* T_*^2}
\label{eq19}
\end{equation}
where $ c_i=n_i\,(c_i=n_i/2) $ is the number of degrees of freedom for boson (fermion) species, and $ \Delta m_i^2 $ is the squared mass difference of particles between two phases. Without massive gauge bosons, the condition $ \alpha>\alpha_{\infty}$ leads to runaway bubbles such that the bubble wall velocity can reach to the speed of light \cite{Bodeker:2009qy}.\footnote[5]{At the next-to-leading order calculation, massive gauge bosons develop an additional frictional term which prevents runaway bubbles \cite{Bodeker:2017cim}.} In this situation, all three sources contribute to the GW spectrum. For $ \alpha<\alpha_{\infty}$, the contribution of bubble collision is negligible and the released energy of the PT is transmitted to the fluid motion. Thus, $ h^2 \Omega(f)\simeq h^2 \Omega_{\mathrm{sw}}(f)+h^2 \Omega_{\mathrm{tu}}(f)$.\\
According to the different values of chemical potential at the critical temperature obtained in the previous section, for different values of $ \lambda_{\mathrm{GB}}$, $ g_*\sim 10$, $ \Delta m^2_i\sim 25\times 10^4$ \cite{BorkaJovanovic:2010yc}, $ N_c=3$, and two heavy quarks, we calculated other key parameters, as listed in Table \ref{t2}.
\begin{table}[H]
	\begin{center}
		\begin{tabular}{| c| c| c| c| c| c|} 
			\hline
			$  N_ f $  & $ \lambda_{\mathrm{GB}}$  & $ T_*[\mathrm{MeV}] $  &  $\alpha$ &$\alpha _{\infty}$& $v_w$ \\
			\hline\hline
			\multirow{7}{*}{2}&0.12  &52.5 &13.4 & 20.7& 0.99\\
			\cline{2-6}
			&0.06 & 104.5 &4.5 & 5.2 & 0.98  \\
			\cline{2-6}
			& 0.03  &132.8 &3.1 & 3.2& 0.97 \\
			\cline{2-6}
			& 0 & 163 & 2.7 &2.1&1 \\ 
			\cline{2-6}
			& -0.03  &192.8 &2.6 &1.5& 1 \\
			\cline{2-6}
			&-0.06  & 221.5 &2.6 &1.16& 1 \\
			\cline{2-6}
			&-0.12  &275.7 &2.56 & 0.7&1\\
			\hline
		\end{tabular}
		\caption{The values of the transition temperature, the ratio of vacuum energy to radiation energy $\alpha$, $\alpha_\infty$, and the bubble wall velocity for $ N_ f= 2$ and different values of Gauss-Bonnet coupling, $ \lambda_{\mathrm{GB}}$, are listed.}
		\label{t2}
	\end{center}
\end{table}
It can be seen from Table \ref{t1}, for negative and zero values of $ \lambda_{\mathrm{GB}}$, we find that $\alpha>\alpha _{\infty}$ and hence $ v_w=1$, while for positive values, $\alpha<\alpha _{\infty}$. In this case, according to the $ \alpha$ values, we use the Jouguet detonation regime in which the bubble wall velocity is given by \cite{Espinosa:2010hh}
\begin{equation}
v_w=\frac{\sqrt{\alpha^2+2\alpha/3}+\sqrt{1/3}}{1+\alpha}.
\end{equation}
As listed in Table \ref{t1}, by enhancing $ \lambda_{\mathrm{GB}}$, $ T_*$ is lowered and $\alpha$ is increased, i.e. it leads to the stronger PT. Also, the effect of increasing the number of heavy quark flavors on $\alpha$ and other parameters can be seen in Appendix \ref{tables}.

\begin{figure}[H]
	\centering
	\includegraphics{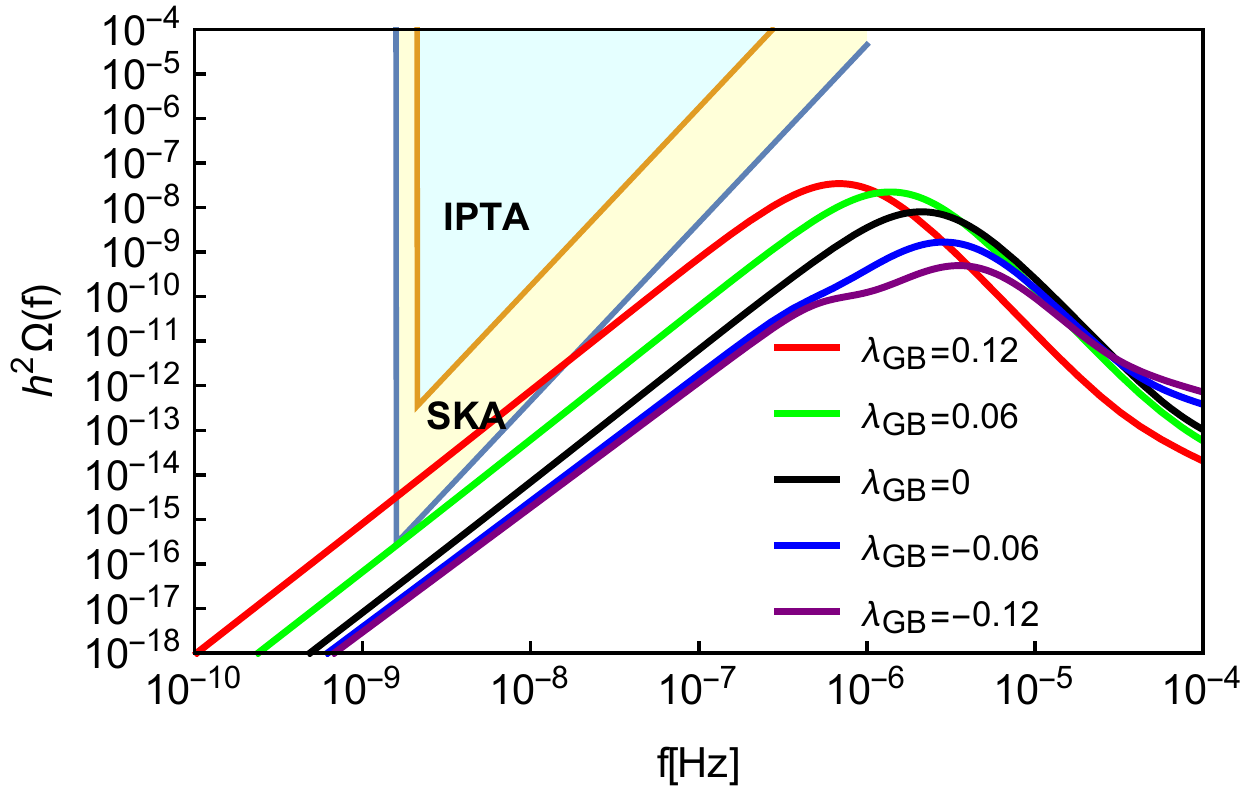}
	\caption{The GW energy density spectrum of the de/confinement PT is displayed for  different values of $\lambda_{\mathrm{GB}}$ and $  N_ f=2 $ case. For $\lambda_{\mathrm{GB}}=0.12$ the GWs might be detected by SKA \cite{Moore:2014lga}. Since for other values of $\lambda_{\mathrm{GB}}$ curves become very close, these curves are not shown.}
	\label{f1}
\end{figure}

By obtaining these key parameters and also quantities pertaining to $\alpha$ and $\alpha_\infty$, see Appendix \ref{gwsources}, we can compute the GW spectrum generated from the first-order de/confinement PT. As can be seen from Fig.\ (\ref{f1}), the GW signals might be detected by SKA telescopes for the case of $ \lambda_{\mathrm{GB}}=0.12$. In particular, the energy density of GW is around $ 10^{-8}$ at the peak frequency, $f_p\sim 10^{-6}\,\mathrm{Hz} $. Moreover, by enhancing $\lambda_{\mathrm{GB}}$, the role of sound wave source to the GW spectrum becomes more important so that, e.g., for $\lambda_{\mathrm{GB}}=-0.03$ sound waves contribute  dominantly to the GW spectrum almost for all frequencies.  

It is also interesting to consider the case in which a period of supercooling occurs at the QCD scales. We study this event and its consequences in the next section.

\subsection{Superrcooling scenario}

In the context of strong first-order PTs, one can investigate the scenario in which the vacuum energy would be very larger compared to the thermal energy of the Universe \cite{Baratella:2018pxi}. In our case with finite baryochemical potential, such a PT might take place and for some e-folds of inflation, the Universe accelerate \cite{Boeckel:2009ej,Jenkovszky:1990ex}. In this section, we take into account a short period of inflation during the QCD PT and explore generated GWs during this stage.

Before the critical temperature, the energy density of the Universe is dominated by the radiation energy. We assume the inflation begins below $ T_c$, the energy density remains constant and hence the Universe expands exponentially. After some e-folds, the radiation part of the energy density becomes negligible compared to the vacuum energy. Subsequently, the energy density decays into the plasma and the inflation can be ended. Therefore, the number of e-folds is calculated as follows
\begin{equation}
N_e=\log\frac{a_e}{a_c}=\log\frac{T_c}{T_e} 	
\end{equation}
where $T_e$ is the exit temperature and $a $ is the scale factor. We approximately consider $ T_e$ as the temperature at which the bubble nucleation and GW production take place.

In the supercooling scenario, $ \alpha$ is given by \cite{Ellis:2019oqb}
\begin{equation}
\alpha\simeq\Big(\frac{\Delta F}{\rho _R}\Big)\Bigg|_{T_e}.	
\end{equation}
Supposing the onset of inflation at $T_c $ and using the energy conservation, we consider the vacuum energy at $T_e $ as $ \Delta F\sim\rho _R(T_c) $. Thus, using $T_c $ values and assuming a short period of inflation, $ N_e=2$, we can find $T_e $ and also $ \alpha$. In addition, with this supercooling stage, we can assume $\beta/H\simeq 10$ \cite{Baratella:2018pxi}.

\begin{figure}[H]
	\centering
	\includegraphics{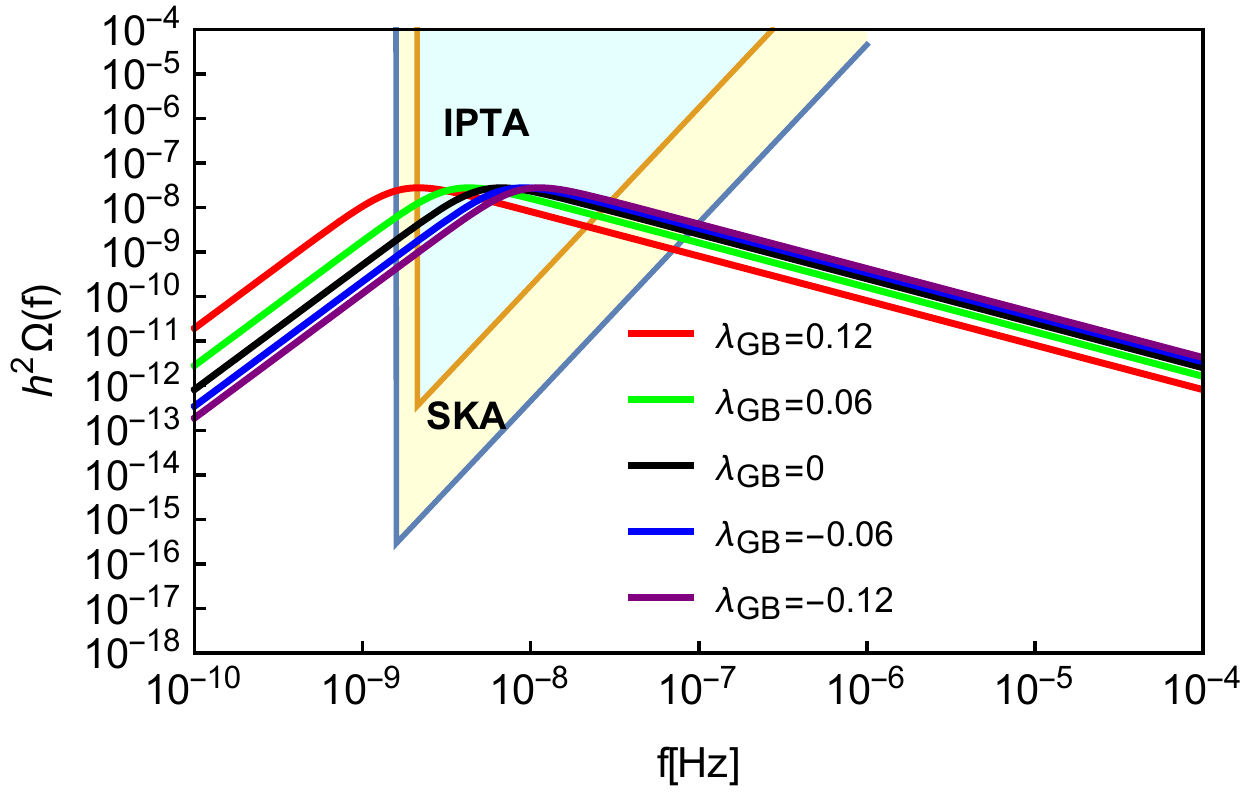}
	\caption{We display the GW energy density spectrum of the de/confinement PT during a short period of supercooling for different values of $\lambda_{\mathrm{GB}}$ and $  N_ f=2 $. The GWs can be detected by SKA and IPTA \cite{Moore:2014lga}.}
	\label{f2}
\end{figure}

In our model for different values of $ \lambda_{\mathrm{GB}}$, $\alpha $ is very large, $\alpha\simeq 2981$. Moreover, due to the inflationary period, the plasma is highly diluted and hence we consider the bubble collision as the main source of the GW generation. As a result, the main fraction of the vacuum energy transfers to the kinetic energy of the bubble wall, $\kappa\simeq 1 $. In this case the bubble wall velocity would be $ v_w\simeq 1$.
\begin{table}[H]
	\begin{center}
		\begin{tabular}{| c| c| c| c|} 
			\hline
			$  N_ f $  & $\lambda_{\mathrm{GB}}$ &$ T_c[\mathrm{MeV}] $ & $ T_e[\mathrm{MeV}] $  \\
			\hline\hline
			\multirow{7}{*}{2}&0.12 &61.7 &8.3\\
			\cline{2-4}
			&0.06  & 123 &16.6  \\
			\cline{2-4}
			& 0.03 &156 &21 \\
			\cline{2-4}
			& 0 & 191.6 & 26\\ 
			\cline{2-4}
			& -0.03  &226.7 &30.7 \\
			\cline{2-4}
			&-0.06 & 260.4 &35.2 \\
			\cline{2-4}
			&-0.12 &324 &43.8 \\
			\hline
		\end{tabular}
		\caption{In the supercooling scenario, the values of the critical temperature and the exit temperature for $ N_ f= 2$ and different values of Gauss-Bonnet coupling, $ \lambda_{\mathrm{GB}}$, are listed.}
		\label{t3}
	\end{center}
\end{table}
As can be seen from Fig.\ (\ref{f2}), in this case the generated GWs can be detected by SKA and IPTA in the near future.

	\section{Conclusion}
	In this paper we have used a soft wall model of AdS/QCD to study GWs produced by the cosmic first-order QCD PT at finite baryochemical potential and coupling. We have considered the effect of leading order inverse coupling corrections perturbatively and the gravity background which includes curvature-squared corrections is known as Gauss-Bonnet gravity. We have also expanded the corrections up to the first-order in $\lambda_{\mathrm{GB}}$.
	
	In our model, the first-order de/confinement PT is corresponded to the first-order Hawking-Page PT in Gauss-Bonnet gravity. We have calculated free energies of the two phases and then explored the critical temperature and baryochemical potential for different values of $\lambda_{\mathrm{GB}}$. Having obtained these parameters, we calculated other key quantities such as the vacuum energy density of the PT and the temperature at which GWs are generated. Using these quantities, we obtained the GW energy density spectrum generated from the first-order de/confinement PT.
	
	Interestingly we find that the GWs might be detected by SKA for sufficiently large Gauss-Bonnet coupling. Moreover, we have found that by enhancing $\lambda_{\mathrm{GB}}$, the role of sound waves in the spectrum of GWs becomes more important. In particular, we have shown for $\lambda_{\mathrm{GB}}<0$, bubbles can runaway so that the bubble wall velocity reaches the speed of light, while for   $\lambda_{\mathrm{GB}}>0$ the wall velocities remains subluminal. In this case, as $\lambda_{\mathrm{GB}}$ is increased, the bubbles move faster. This result is also consistent with the fact that the higher $\lambda_{\mathrm{GB}}$, the lower viscosity.
	
	Furthermore, we have considered a supercooling scenario, during the QCD phase transition, where the vacuum energy could be very larger than the thermal energy of the Universe. The plasma is highly diluted in this case and we considered only the bubble collision as the main source of GWs. We have obtained the vacuum energy to thermal energy density ratio and shown that it is very large for different vales of $\lambda_{\mathrm{GB}}$. We also find that the GWs generated during this period can be detected by SKA and IPTA experiments. 

\begin{appendices}
	\section{Gravitational wave sources}\label{gwsources}
	In this appendix, we review the contribution of three sources to the energy density of GWs generated during a first-order PT.\\
	Using the envelope approximation, the contribution of the bubble collision to the GW frequency spectrum is calculated by numerical simulations as \cite{Kamionkowski:1993fg}
	\begin{equation}
	h^2\Omega _{\mathrm{col}}(f)=1.67\times 10^{-5}\Big(\frac{0.11 v_{w}^3}{0.42+v_{w}^2}\Big) \Big(\frac{H_*}{\beta}\Big)^{2}\Big(\frac{\kappa \alpha}{1+\alpha}\Big)^2 \Big(\frac{100}{g_*}\Big)^{\frac{1}{3}}S_{\mathrm{en}}(f)
	\end{equation}
	where $ \kappa$ denotes the fraction of the vacuum energy converted into the kinetic energy of the bubbles. Also, the spectral shape of the GW is obtained by the following analytic fit \cite{Huber:2008hg}
	\begin{equation}
	S_{\mathrm{en}}(f)=\frac{3.8(\frac{f}{f_{\mathrm{en}}})^{2.8}}{1+2.8(\frac{f}{f_{\mathrm{en}}})^{3.8}}
	\end{equation}
	where the present red-shifted peak frequency is given by
	\begin{equation}
	f_{\mathrm{en}}=16.5\times 10^{-6}[\mathrm{Hz}] \Big(\frac{0.62}{1.8-0.1 v_{w}+v_{w}^2}\Big)\Big(\frac{\beta}{H_*}\Big)\Big(\frac{T_*}{100~\mathrm{GeV}}\Big)\Big(\frac{g_*}{100}\Big)^{\frac{1}{6}}.
	\end{equation}
	The GW contribution form sound waves is numerically calculated by \cite{Hindmarsh:2015qta}
	\begin{equation}
	h^2\Omega _{\mathrm{sw}}(f)=2.65\times 10^{-6}\Big(\frac{H_*}{\beta}\Big)\Big(\frac{\kappa _{\mathrm{sw}} \alpha}{1+\alpha}\Big)^2\Big(\frac{100}{g_*}\Big)^{\frac{1}{3}} v_{w}~ S_{\mathrm{sw}}(f)
	\end{equation}
	where the spectral shape is given by \cite{Caprini:2015zlo}
	\begin{equation}
	S_{\mathrm{sw}}(f)=\Big(\frac{f}{f_{\mathrm{sw}}}\Big)^3\Big(\frac{7}{4+3(\frac{f}{f_{\mathrm{sw}}})^{2}}\Big)^{\frac{7}{2}}
	\end{equation}
	and its red-shifted peak frequency is
	\begin{equation}
	f_{\mathrm{sw}}=1.9\times 10^{-5}[\mathrm{Hz}]  \Big(\frac{1}{v_{w}}\Big)\Big(\frac{\beta}{H_*}\Big)\Big(\frac{T_*}{100~\mathrm{GeV}}\Big)\Big(\frac{g_*}{100}\Big)^{\frac{1}{6}}.
	\end{equation}
	Furthermore, the contribution of the turbulent motion of the fluid is given by
	\begin{equation}
	h^2\Omega _{\mathrm{tu}}(f)=3.35\times 10^{-4}\Big(\frac{H_*}{\beta}\Big)\Big(\frac{\kappa _{\mathrm{tu}} \alpha}{1+\alpha}\Big)^{\frac{3}{2}}\Big(\frac{100}{g_*}\Big)^{\frac{1}{3}} v_{w}~ S_{\mathrm{tu}}(f)
	\end{equation}
	where the spectral shapes is as follows \cite{Caprini:2015zlo},
	\begin{equation}
	S_{\mathrm{tu}}(f)=\frac{(\frac{f}{f_{\mathrm{tu}}})^3}{(1+\frac{f}{f_{\mathrm{tu}}})^{\frac{11}{3}} (1+\frac{8\pi f}{h_*})}
	\end{equation}
	with
	\begin{equation}
	h_*=16.5\times 10^{-6} [\mathrm{Hz}]\Big(\frac{T_*}{100~\mathrm{GeV}}\Big)\Big(\frac{g_*}{100}\Big)^{\frac{1}{6}}
	\end{equation}
	as the red-shifted Hubble parameter. The red-shifted peak frequency is given by
	\begin{eqnarray}
	f_{\mathrm{tu}}=2.7\times 10^{-5}[\mathrm{Hz}] \Big(\frac{1}{v_{w}}\Big)\Big(\frac{\beta}{H_*}\Big)\Big(\frac{T_*}{100~\mathrm{GeV}}\Big)\Big(\frac{g_*}{100}\Big)^{\frac{1}{6}}.
	\end{eqnarray}
	For runaway bubbles, the efficiency factor of different sources is expressed as follows. For the bubble collision source, the efficiency factor is given by
	\begin{equation}
	\kappa=1-\frac{\alpha_{\infty}}{\alpha}
	\end{equation}
	where $ \alpha_{\infty}/\alpha$ is the fraction transformed into the fluid motion and thermal energy. In addition, the fraction which is converted to the plasma motion is given by
	\begin{equation}
	\kappa_v=\frac{\alpha_{\infty}}{\alpha}\frac{\alpha_{\infty}}{0.73+0.083\sqrt{\alpha_{\infty}}+\alpha_{\infty}}
	\end{equation}
	where the contribution of sound waves is expressed as $ \kappa_{\mathrm{sw}}=(1-\delta)\kappa_v $ and the fraction of plasma motion which is turbulence, $ \delta=\kappa_{\mathrm{tu}}/\kappa_v$, can be of the order of $\delta=0.1 $ \cite{Caprini:2015zlo}.
	
	For non-runaway bubbles, $ \kappa_v$ is expressed as
	\begin{equation}
	\kappa_v=\frac{\alpha}{0.73+0.083\sqrt{\alpha}+\alpha}.
	\end{equation}
	\section{Tables of phase transition parameters}\label{tables}
	In this appendix, the effect of changing the number of heavy quark flavors on important characteristics of the PT is studied.
	\begin{table}[H]
		\begin{center}
			\begin{tabular}{| c| c| c| c| c| c| c| c|} 
				\hline
				$  N_ f $  & $\lambda_{\mathrm{GB}}$  & $ \mu [\mathrm{MeV}]$&$ T_c[\mathrm{MeV}] $ & $ T_*[\mathrm{MeV}] $ & $\alpha$ &$\alpha _{\infty}$&$v_w$ \\
				\hline\hline
				\multirow{7}{*}{1} & 0.12 &50 &61.9 &52.7 & 13.5 &20.5&0.99 \\
				\cline{2-8}
				& 0.06 &102 &122.9 &104.6 &4.5 &5.2&0.98 \\
				\cline{2-8}
				& 0.03 &113 &156.1&132.8&3.1 &3.2&0.97 \\
				\cline{2-8}
				& 0 &121& 191.6& 163 & 2.7 &2.1 &1\\ 
				\cline{2-8}
				& -0.03 & 128 &226.6 &192.8 &  2.6 &1.5&1 \\
				\cline{2-8}
				&-0.06 &134 & 260.4 & 221.5 &2.6&1.1 &1 \\
				\cline{2-8}
				&-0.12 &144 &324 &275.6 & 2.5 & 0.7&1\\
				\hline
			\end{tabular}
			\caption{The values of the chemical potential, the critical temperature, the transition temperature, the ratio of vacuum energy to radiation energy, $\alpha$, and $\alpha_\infty$ for $ N_ f= 1$ and different values of Gauss-Bonnet coupling, $ \lambda_{\mathrm{GB}}$, are listed.}
			\label{t4}
		\end{center}
	\end{table}
	
	\begin{table}[H]
		\begin{center}
			\begin{tabular}{| c| c| c| c| c| c| c| c|} 
				\hline
				$  N_ f $  & $\lambda_{\mathrm{GB}}$  & $ \mu [\mathrm{MeV}]$&$ T_c[\mathrm{MeV}] $ & $ T_*[\mathrm{MeV}] $  & $\alpha$ &$\alpha _{\infty}$&$v_w$ \\
				\hline\hline
				\multirow{7}{*}{3} & 0.12 &28 & 61.2 &52.1 & 13.1 &20.9&0.99 \\
				\cline{2-8}
				& 0.06 & 59 & 122.9 &104.6 &4.5 &5.2&0.98 \\
				\cline{2-8}
				& 0.03 &65 &156.1 &132.8 &2.6 &3.2&0.97 \\
				\cline{2-8}
				& 0 &70& 191.6& 162.9 & 2.7 &2.1&1 \\ 
				\cline{2-8}
				&- 0.03 &74 &226.6 &192.8 & 2.6 &1.5&1 \\
				\cline{2-8}
				&-0.06 & 77 & 260.5 &221.6 &2.6 &1.1&1  \\
				\cline{2-8}
				&-0.12 &83 &324 &275.6 &2.5 & 0.7&1\\
				\hline
			\end{tabular}
			\caption{The values of the chemical potential, the critical temperature, the transition temperature, the ratio of vacuum energy to radiation energy, $\alpha$, and $\alpha_\infty$ for $ N_ f= 3$ and different values of Gauss-Bonnet coupling, $ \lambda_{\mathrm{GB}}$, are listed.} 
			\label{t5}
		\end{center}
	\end{table}
	
\end{appendices}


\end{document}